\documentclass{svproc}
\usepackage{url}

\usepackage{graphicx}
\usepackage[rightcaption]{sidecap}

\begin{document}
\mainmatter

\title{Extracting Network Structures from Corporate Organization Charts Using Heuristic Image Processing}
\titlerunning{Extracting Network Structures from Corporate Organization Charts}

\author{Hiroki Sayama\inst{1,2} and Junichi Yamanoi\inst{2}}
\authorrunning{H. Sayama and J. Yamanoi}

\institute{Binghamton Center of Complex Systems, Binghamton University,\\State University of New York, Binghamton, NY 13902-6000, USA \and
Waseda Innovation Lab, Waseda University, Shinjuku, Tokyo 169-8050, Japan
\email{sayama@binghamton.edu}, \email{yamanoi@waseda.jp}}
\maketitle              
\begin{abstract}
Organizational structure of corporations has potential to provide implications for dynamics and performance of corporate operations. However, this subject has remained unexplored because of the lack of readily available organization network datasets. To overcome the this gap, we developed a new heuristic image-processing method to extract and reconstruct organization network data from published organization charts. Our method analyzes a PDF file of a corporate organization chart and detects text labels, boxes, connecting lines, and other objects through multiple steps of heuristically implemented image processing. The detected components are reorganized together into a Python's NetworkX Graph object for visualization, validation and further network analysis. We applied the developed method to the organization charts of all the listed firms in Japan shown in the ``Organization Chart/System Diagram Handbook'' published by Diamond, Inc., from 2008 to 2011. Out of the 10,008 organization chart PDF files, our method was able to reconstruct 4,606 organization networks (data acquisition success rate: 46\%). For each reconstructed organization network, we measured several network diagnostics, which will be used for further statistical analysis to investigate their potential correlations with corporate behavior and performance.
\end{abstract}

\section{Introduction}

Formal organizational structure of corporations, which represents official relationships among members and formal pathways of corporate chains of command, has potential to provide implications for dynamics and performance of corporate operations \cite{mintzberg1979,argyres2004,reitzig2022}. However, this subject has not been fully explored computationally in the fields of organization science, management science, social network analysis, data science, and complex systems science, primarily because of the lack of readily available organization network datasets. Most of the information about corporate organizational structures are publicized in a graphical organization chart, which may be used for a small number of manual case studies but would not be suitable for large-scale quantitative statistical analysis. Because of this limitation in data availability, earlier studies typically used rather simple indirect information only (e.g., short description of management style \cite{csaszar2012}, survey responses \cite{gentile2020}) as a surrogate variable to represent organizational structure. Research on more detailed topological properties of organizational structure thus remains scarce and fragmented \cite{reitzig2015,joseph2020}.

To overcome the above gap in corporate organization research, we developed a new heuristic image-processing method to extract and reconstruct organization network data from published organization charts. We expect that such computer-readable network datasets will be quite useful for investigations of corporate organizations, especially for seeking potential correlations between detailed organization network properties and corporate dynamics/performance. We also hope that this work presents a counterproposition to the current Artificial Intelligence (AI) / Machine Learning (ML) mainstream trends by providing a concrete example of what one can do when training data is not available for the problem to be solved. 

\section{Dataset}

\begin{figure}[tp]
\centering
\includegraphics[width=0.325\textwidth]{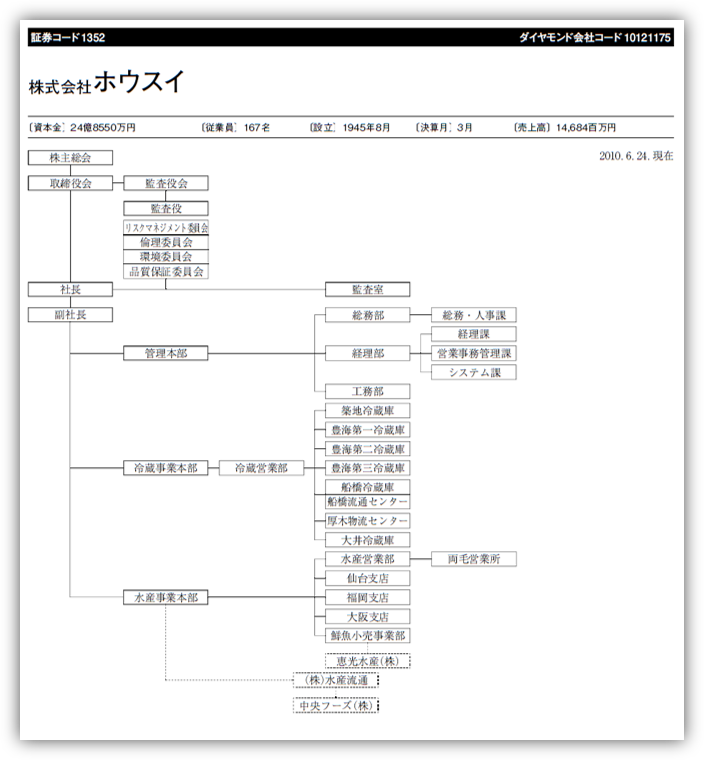}
\includegraphics[width=0.325\textwidth]{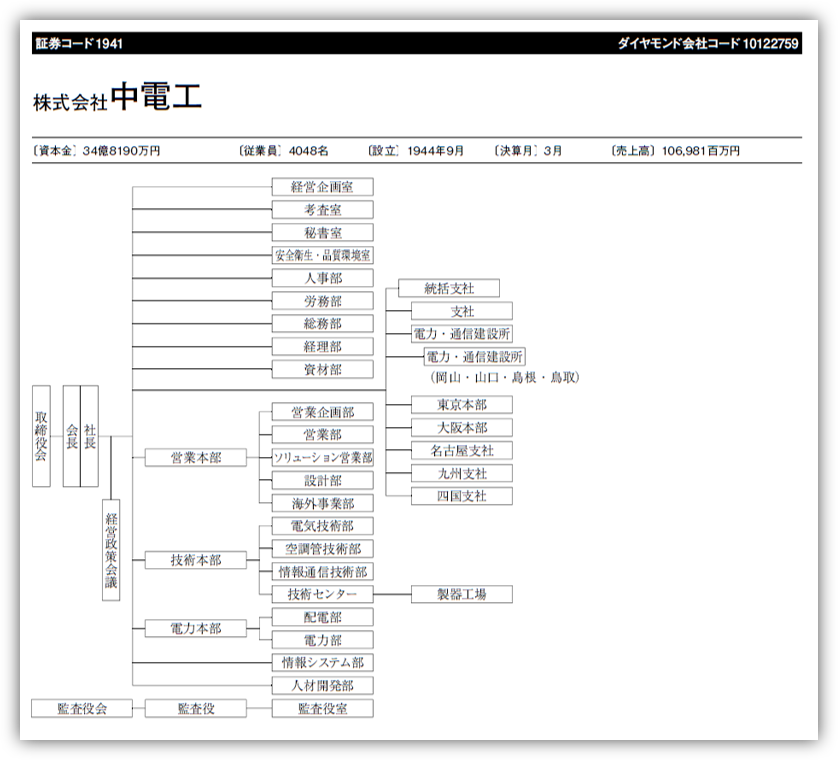}
\includegraphics[width=0.325\textwidth]{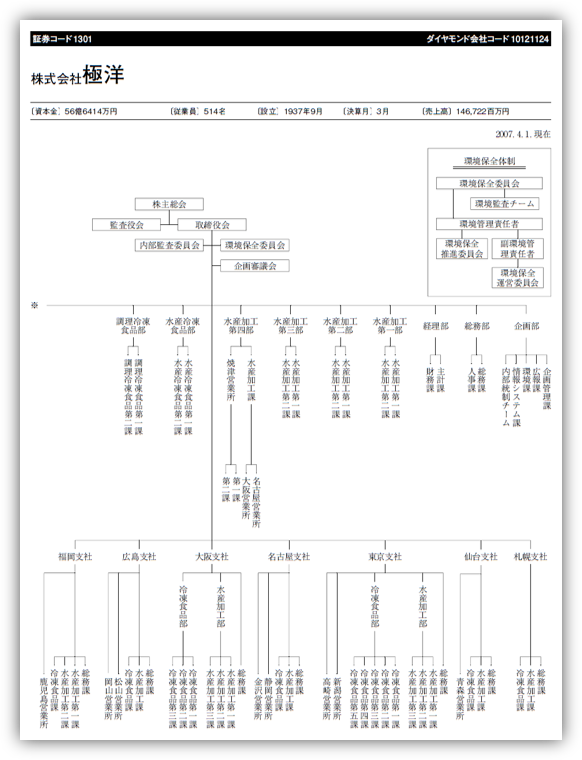}
\caption{Examples of organization charts published in the data source \cite{diamond}.}
\label{examples}
\end{figure}

The data we aim to extract network structures from are the organization charts of all the listed firms in Japan from 2008 to 2011, which were provided in the ``Organization Chart/System Diagram Handbooks'' (original Japanese title: ``Soshikizu-Keitouzu Binran'') published by Diamond, Inc. These books (and their electronic CD-ROM versions) included a total of 10,008 organization chart PDF files. Those charts were apparently drawn manually in a free format without a strict template or annotations. Some examples of the organization charts are shown in Fig.~\ref{examples}. Each of these charts was provided simply as a single PDF file in the dataset with no additional computer-readable annotations about its organization network structure.

\section{Method}

The currently popular methods for information extraction from images are AI/ ML-based ones, typically involving convolutional neural networks (CNNs). Such methods, however, would not work in problems like the one we are trying to solve in this study, because of the lack of training datasets. In order to apply training-based AI/ML methods, we would need to have a large number of pairs of organization chart images and their annotated network representations. This limitation suggests that we need to take a different, more logic-based, manually designed, heuristic approach.

In this study, we designed and implemented a fully automated method that can analyze a PDF file of a corporate organization chart and detect text labels, boxes, connecting lines, and other objects through 10 steps of image processing. The application was written in Python 3 with several packages listed in Table \ref{packages}. The detected components are reorganized together into a Python's NetworkX Graph object for visualization, validation and further network analysis. Details of the 10 steps in the developed method are described below.
\begin{table}[tp]
\centering
\caption{Packages used in the implementation of the developed network structure extraction method.}
\label{packages}
\begin{tabular}{ll}
\hline
Name & Purpose \\
\hline
pdf2image & To convert a PDF to a bitmap image \\
PIL & For image manipulation \\
OpenCV & For image manipulation \\
NetworkX & For network representation, visualization and analysis \\
pdfminer & To extract text labels within a PDF\\
\hline
\end{tabular}
\end{table}

\begin{SCfigure}[][tp]
\centering
\includegraphics[scale=0.2]{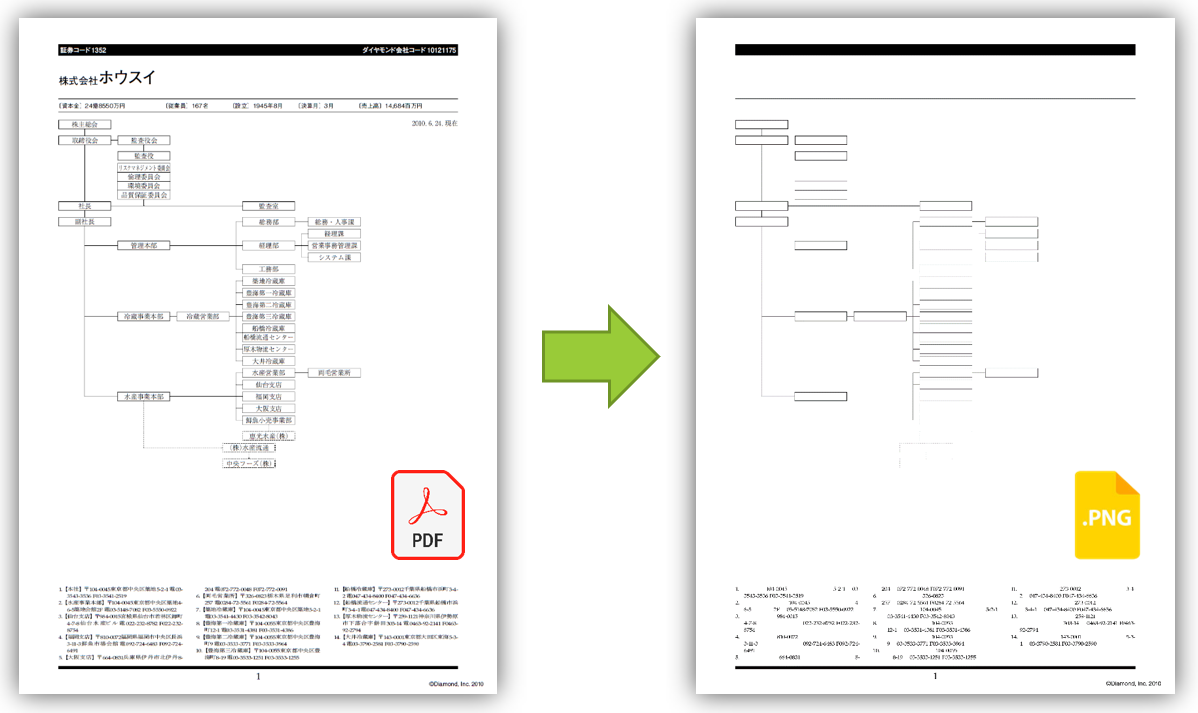}
\caption{Step 1: Converting the original PDF to a PNG image format and removing Japanese texts from the chart.}
\label{step1}
\end{SCfigure}

The first step is to convert the original PDF file into a bitmap image format (PNG) using pdf2image (Fig.~\ref{step1}). This step automatically removes all Japanese characters from the image since the Japanese character set was not supported by pdf2image, which was a convenient property for further image analysis steps.

\begin{SCfigure}[][tp]
\centering
\includegraphics[scale=0.2]{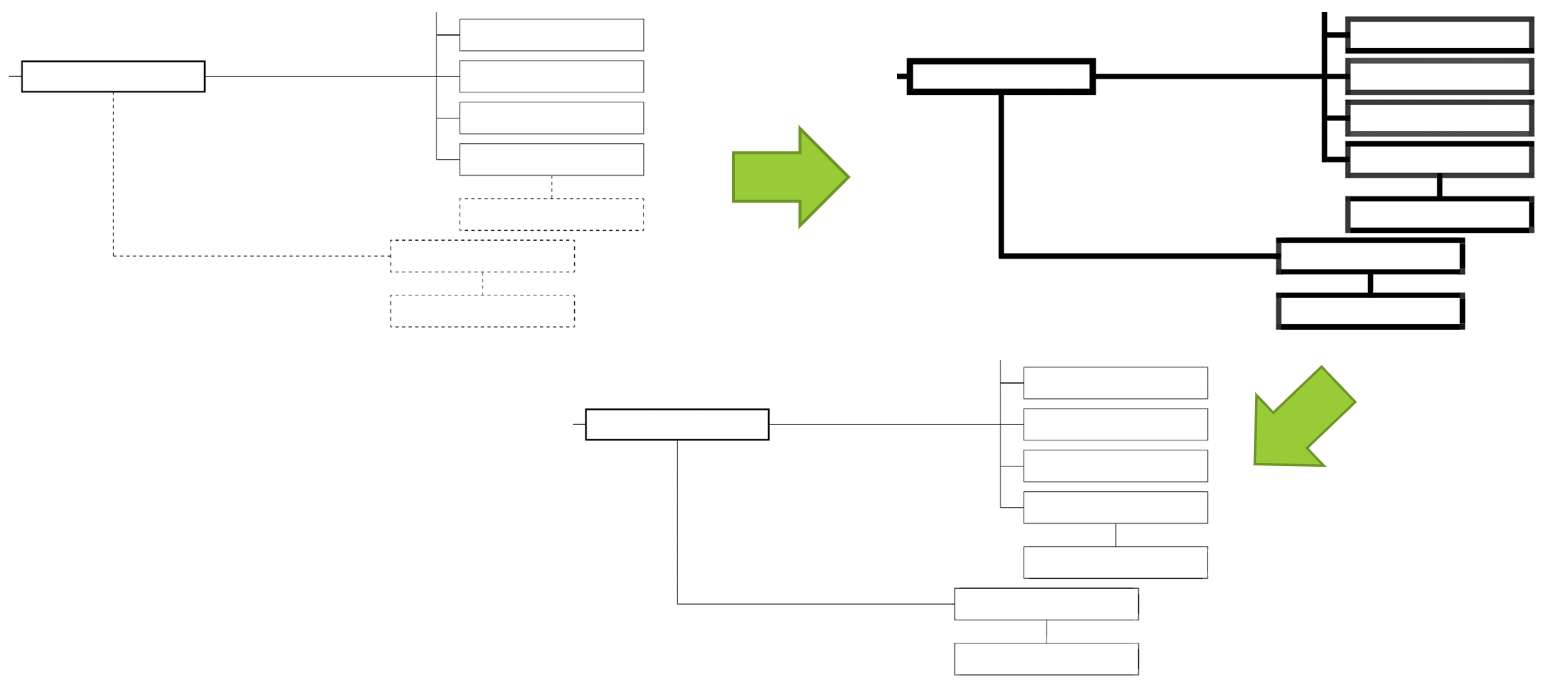}
\caption{Step 2: Connecting dashed/dotted lines into solid lines through image dilation and erosion operations.}
\label{step2}
\end{SCfigure}

The second step is to use image dilation and erosion operations to connect dashed/dotted lines into solid lines (Fig.~\ref{step2}). This step was necessary because many organization charts used dashed/dotted lines to represent connections and boxes.

\begin{SCfigure}[][tp]
\centering
\includegraphics[scale=0.2]{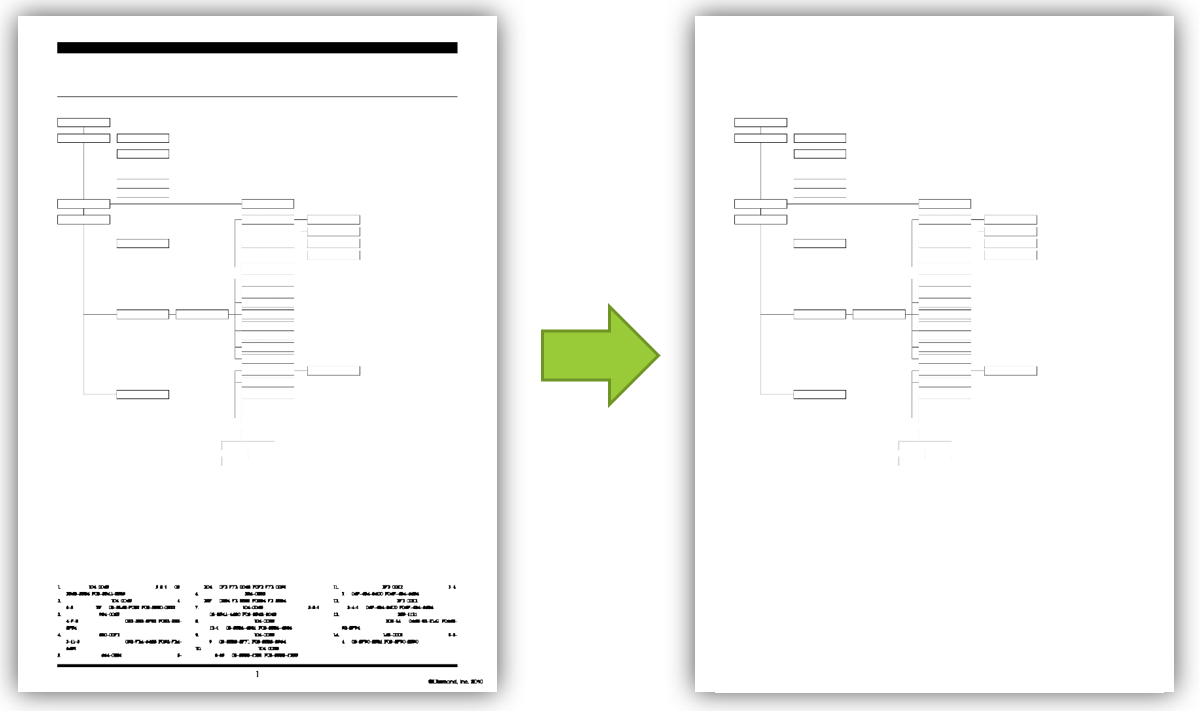}
\caption{Step 3: Removing header and footer areas.}
\label{step3}
\end{SCfigure}

The third step is to remove header and footer areas that do not contain organization chart information (Fig.~\ref{step3}). The header size was the same for all the charts in this dataset, so we simply erased any content in the top portion of each image. Meanwhile, the footer size varied from chart to chart, so we scanned the image to detect a major separation gap (i.e., a substantial number of blank horizontal rows in the image) between the chart and the footer and erased any content beneath it. After this step, the resulting image contains only a wire diagram showing the organizational structure (Fig.~\ref{step3}, right).

\begin{SCfigure}[][tp]
\centering
\includegraphics[scale=0.2]{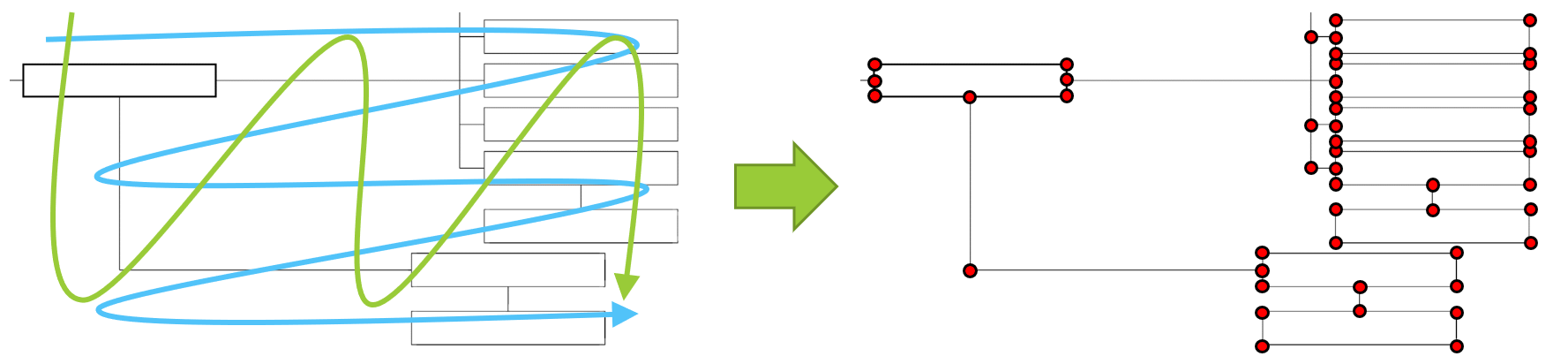}
\caption{Step 4: Detecting all the horizontal and vertical line segments as edges by scanning the entire image in two ways.}
\label{step4}
\end{SCfigure}

The fourth step is to detect all the horizontal and vertical line segments as edges in the image. This was done by scanning the image twice, horizontally and vertically, and detecting all consecutive sequences of black pixels as edges (Fig.~\ref{step4}). In this step, physically very close nodes were aggregated into a single node, and any accidentally created self-loops were removed. These corrections were repeated multiple times until no more such cases were present in the edges.

\begin{SCfigure}[][tp]
\centering
\includegraphics[scale=0.2]{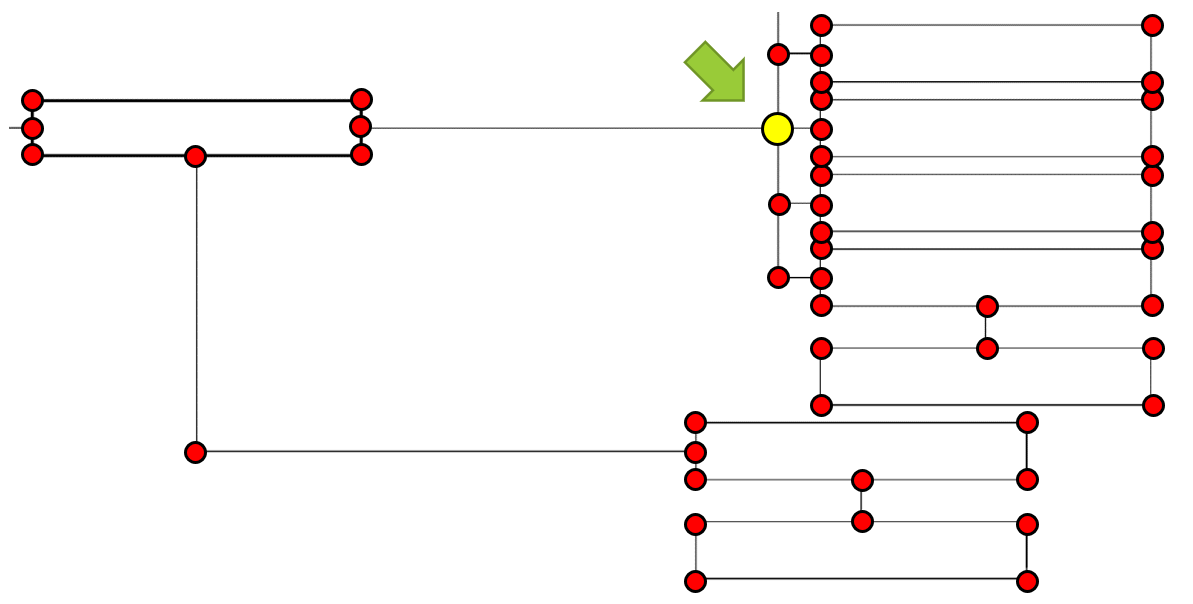}
\caption{Step 5: Inserting new nodes to intersection points of horizontal and vertical edges.}
\label{step5}
\end{SCfigure}

The fifth step is to insert new nodes to wherever horizontal and vertical edges intersect (Fig.~\ref{step5}).

\begin{SCfigure}[][tp]
\centering
\includegraphics[scale=0.2]{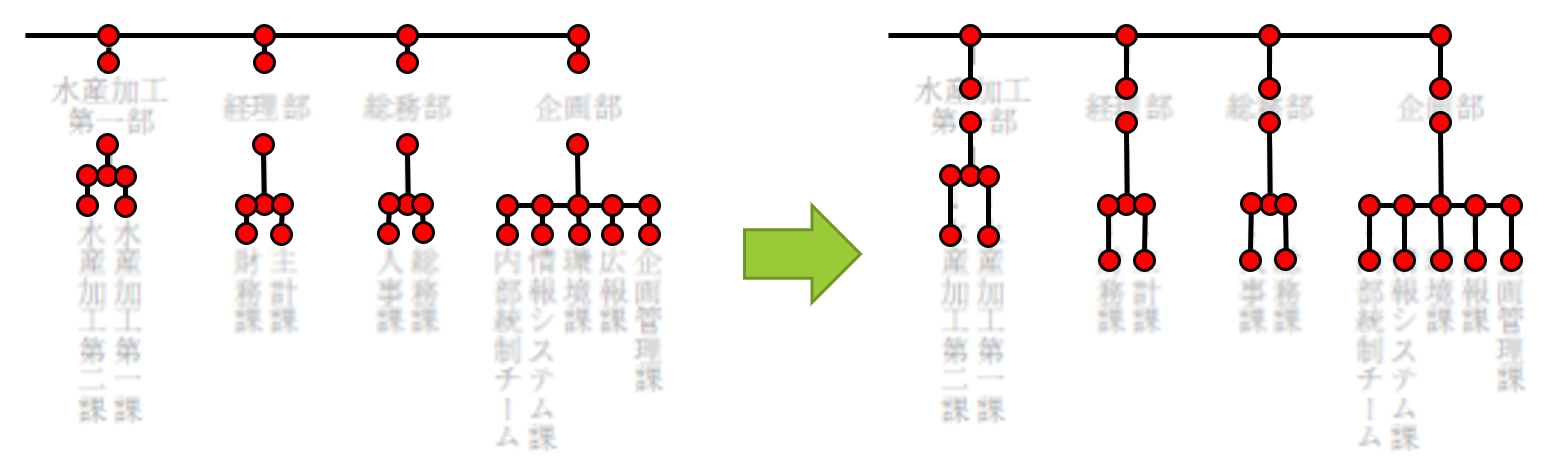}
\caption{Step 6: Extending edges incidental to degree-1 nodes a little outwards.}
\label{step6}
\end{SCfigure}

The six step is to extend edges incidental to degree-1 nodes a little outwards (Fig.~\ref{step6}). This step was needed in order to make the node-textbox matching (discussed later) more robust, especially for unboxed text labels.

\begin{SCfigure}[][tp]
\centering
\includegraphics[scale=0.2]{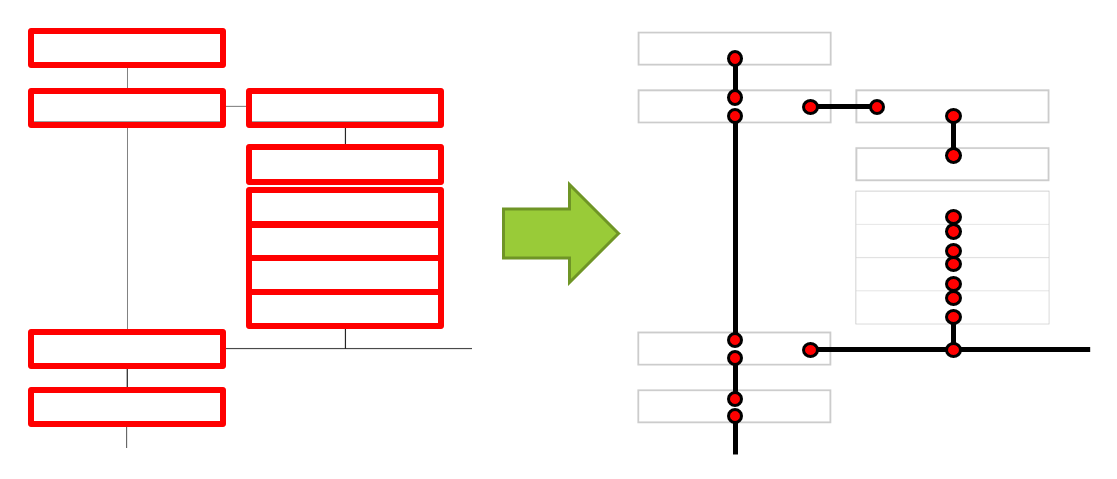}
\caption{Step 7: Detecting boxes (minimal cycles in the cycle basis) and replacing them with single nodes appropriately.}
\label{step7}
\end{SCfigure}

The seventh step is to detect boxes (minimal cycles in the cycle basis) and replace them with single nodes appropriately (Fig.~\ref{step7}). This step was needed because edges in such boxes were not about organizational structure but just for visual borders of boxes. More details of this step are as follows:
\begin{itemize}
\item An external edge connecting to a rectangular circle were connected directly to the inside area of the rectangle.
\item Two rectangles with a shared edge are replaced by a new edge connecting the two areas of the rectangles.
\item Rectangles with no outside connection are represented as an isolated node.
\item After all of the above, all minimal remaining cycles are removed.
\end{itemize}

\begin{SCfigure}[][tp]
\centering
\includegraphics[scale=0.2]{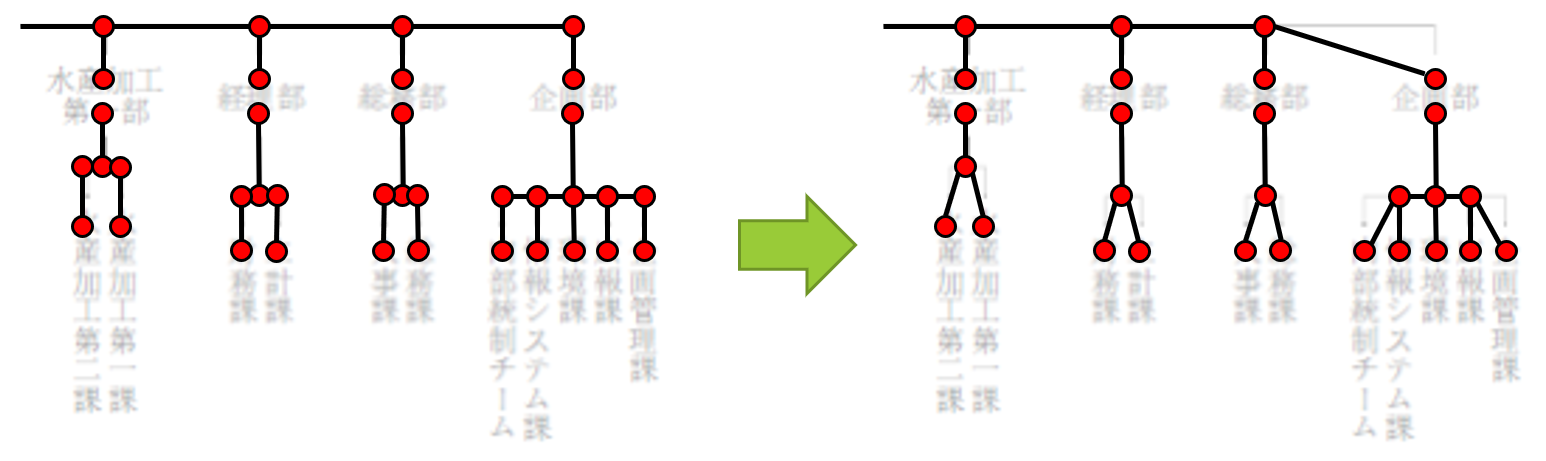}
\caption{Step 8: Removing degree-2 nodes and connecting their neighbors directly.}
\label{step8}
\end{SCfigure}

The eighth step is to remove all degree-2 nodes and connect their neighbors directly (Fig.~\ref{step8}). This was needed because such nodes were topologically unnecessary in organization network structure.

\begin{SCfigure}[][tp]
\centering
\includegraphics[scale=0.2]{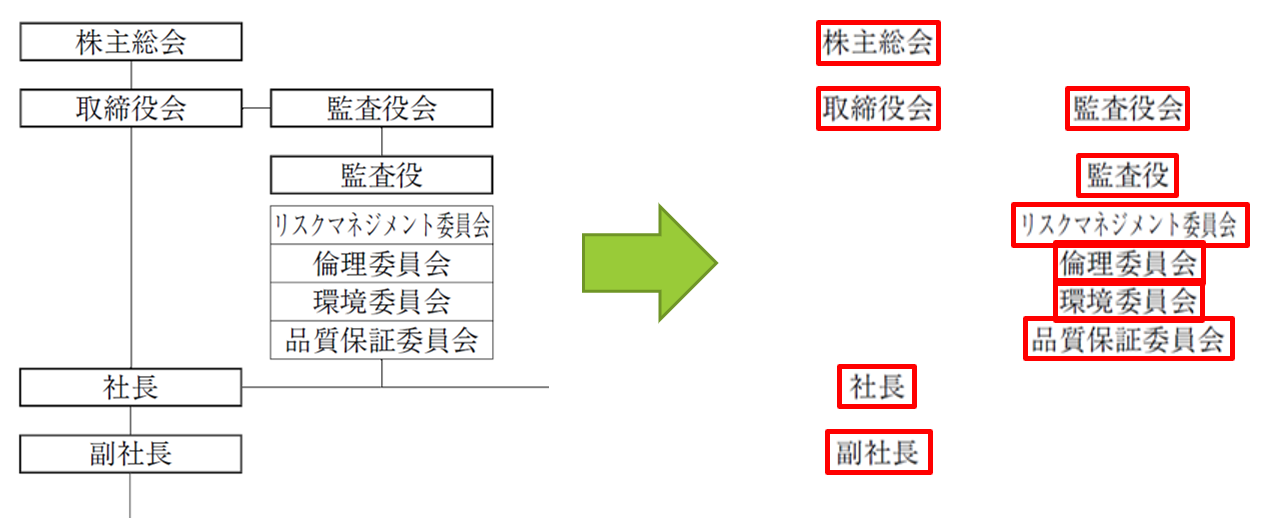}
\caption{Step 9: Detecting textboxes in the original PDF and recording their content, position and size.}
\label{step9}
\end{SCfigure}

The ninth step is to detect all the textboxes in the original PDF and record their content, position and size (Fig.~\ref{step9}). Some textboxes included multiple lines containing multiple node labels, which were automatically split into several textboxes.

\begin{SCfigure}[][tp]
\centering
\includegraphics[scale=0.2]{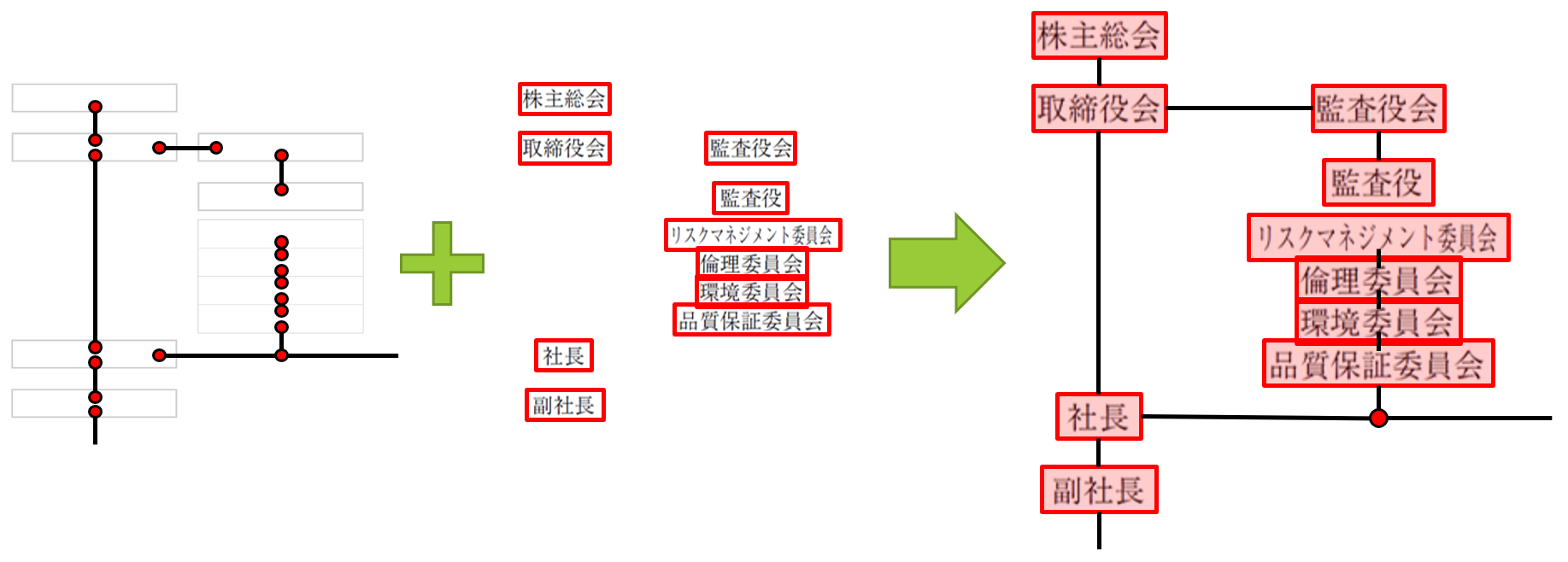}
\caption{Step 10: Merging network nodes and textboxes by matching their positions.}
\label{step10}
\end{SCfigure}

Finally, the tenth step is to merge the detected network nodes and the textboxes according to their positions (Fig.~\ref{step10}). This results in a single network representation of the original organization chart.

\section{Results}

\begin{figure}[tp]
\centering
\includegraphics[height=0.495\textwidth]{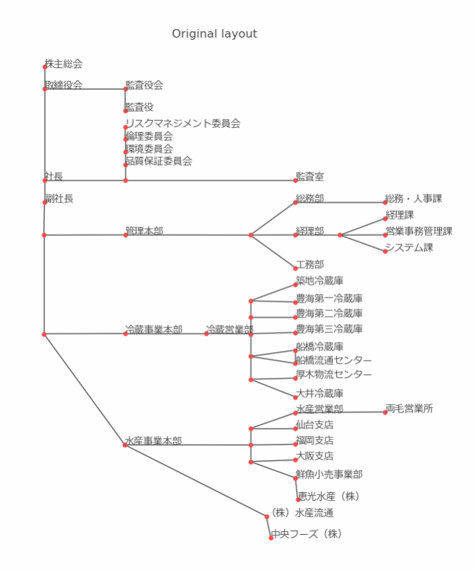}
\includegraphics[width=0.495\textwidth]{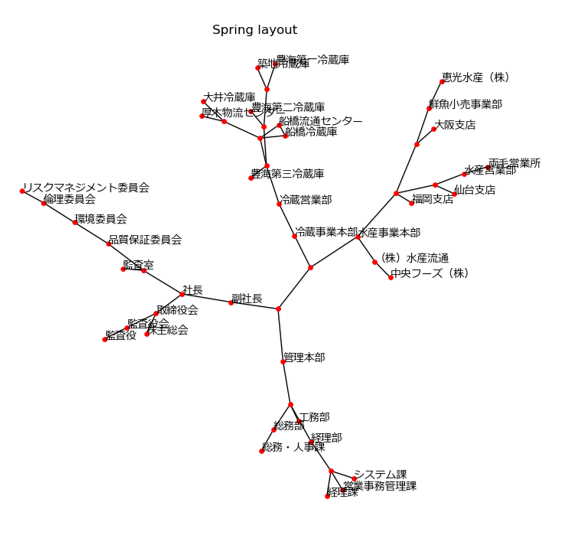}\\
~\\
\includegraphics[height=0.495\textwidth]{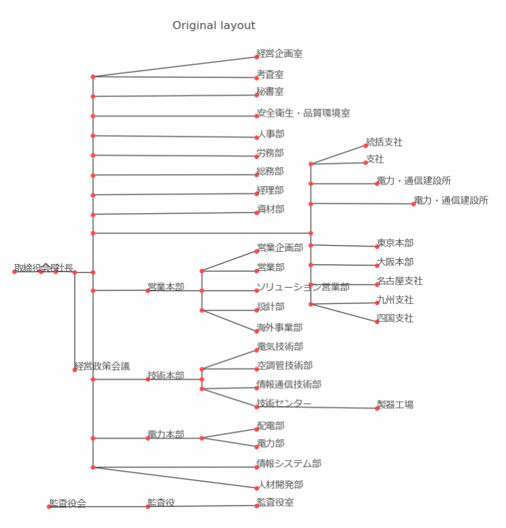}
\includegraphics[width=0.495\textwidth]{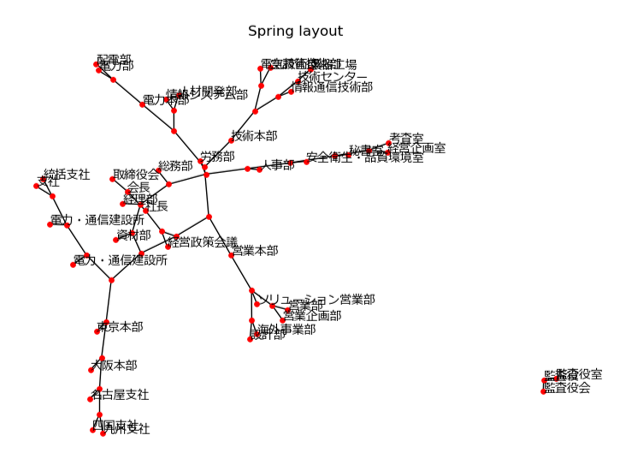}\\
~\\
\includegraphics[width=0.495\textwidth]{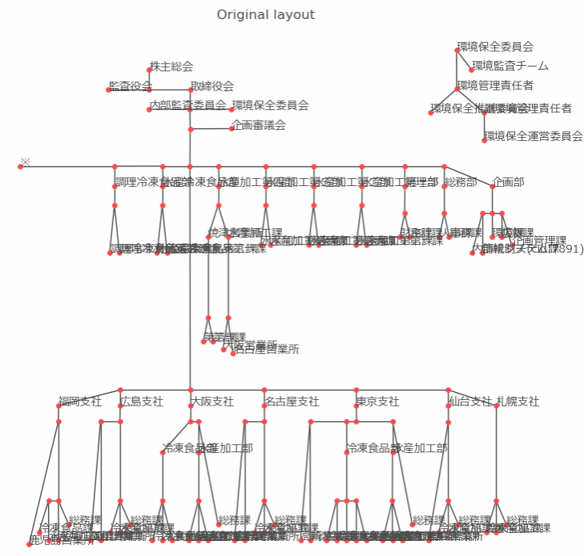}
\includegraphics[width=0.495\textwidth]{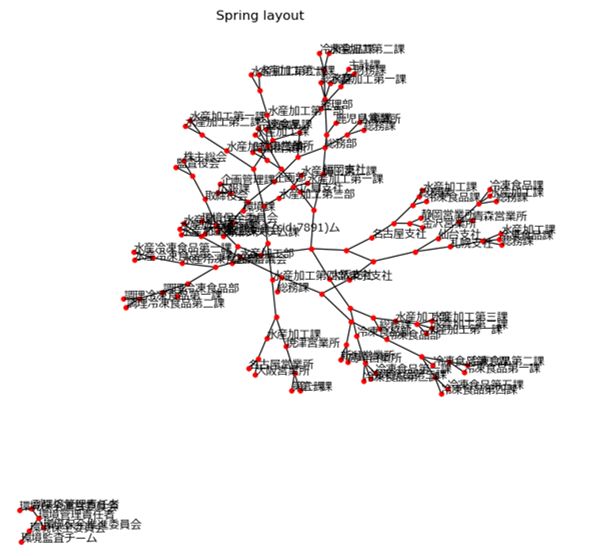}
\caption{Examples of network extraction results. The results for the three examples in Fig.~\ref{examples} are shown. Left column: Networks drawn following original chart layouts. Right: Networks drawn using NetworkX's spring layout algorithms.}
\label{results}
\end{figure}

We applied the developed method to all the organization charts published in \cite{diamond}. Many of the charts did not follow typical organizational diagram format and thus our method failed on them. However, out of the 10,008 organization chart PDF files, our method was able to reconstruct 4,606 organization networks, making the data acquisition success rate to be 46\%. While there are no other methods to compare with, we consider this a very high success rate, given that the method was designed and implemented entirely based on logic and heuristics, without using any AI/ML tools. Some of the extracted networks are shown in Fig.~\ref{results}.

For each reconstructed organization network, we measured several network diagnostics, including number of nodes, number of edges, density, average clustering coefficient, number of connected components, average shortest path length, and algebraic connectivity. We are currently investigating potential correlations between those network diagnostics and corporate behavior and performance, which will be reported elsewhere \cite{inpreparation}.

\section{Conclusions}

Here we reported the algorithmic details of the heuristic image-processing method we developed for automated extraction of network structures from organization charts. It achieved a close-to-50\% success rate of data acquisition on actual organization charts of listed firms in Japan, which serves as a proof that such logic-based approaches for tool development and data analysis can be useful when training data are missing or unavailable at all. We note that our method is still limited as it detects only horizontal/vertical lines as edges, and it was tested only with Japanese organization charts published in \cite{diamond}. Similar tools should be developed and tested for more complicated, multilingual charts in the future.

\section*{Acknowledgments}

We thank Dr.~Samiksha Shukla at Christ University for her continuous encouragement and support for preparation of this manuscript.

\end{document}